\documentclass[aps,pra,reprint]{revtex4-1}
\usepackage{graphicx}
\usepackage{amsmath}
\usepackage{amsfonts}

\usepackage{hyperref}
\hypersetup{
colorlinks=true, 
citecolor=black, 
linkcolor=black}

\newcommand{\fref}[1]{Fig.~\ref{#1}}

\newcommand{\pref}[1]{(\ref{#1})}

\newcommand{\eref}[1]{Eq.~\pref{#1}}

\begin{document}

\title{Tunneling-induced angular momentum for single cold atoms}
\author{R. Menchon-Enrich,$^1$ S. McEndoo,$^2$ J. Mompart,$^1$ V. Ahufinger,$^1$ and Th. Busch$^{3,4}$}

\affiliation{$^1$Departament de F\'{\i}sica, Universitat Aut\`{o}noma de Barcelona, E-08193 Bellaterra, Spain } 
\affiliation{$^2$SUPA, EPS/Physics, Heriot--Watt University, Edinburgh, EH14 4AS, United Kingdom} 
\affiliation{$^3$Physics Department, University College Cork, Cork, Ireland}
\affiliation{$^4$Quantum Systems Unit, OIST Graduate University, Okinawa 904-0495, Japan} 
\date{\today}
\begin{abstract}
We study the generation of angular momentum carrying states for a single cold particle by breaking the symmetry of a spatial adiabatic passage process in a two-dimensional system consisting of three harmonic potential wells. By following a superposition of two eigenstates of the system, a single cold particle is completely transferred to the degenerate first excited states of the final trap, which are resonantly coupled via tunneling to the ground states of the initial and middle traps. Depending on the total time of the process, angular momentum is generated in the final trap, with values that oscillate between $\pm\hbar$. This process is discussed in terms of the asymptotic eigenstates of the individual wells and the results have been checked by simulations of the full two-dimensional Schr\"odinger equation.
\end{abstract}
\pacs{PACS}

\maketitle
\section{Introduction}
Controlling the states of quantum particles is a challenging task and a topic of significant present activity in the fields of atom optics, quantum computation, quantum metrology, and quantum simulation of condensed matter systems \cite{general}. In particular, the generation of angular momentum for matter waves is attracting a lot of attention as, for instance, in studying superfluid properties of Bose-Einstein condensates (BECs) \cite{superfluidity}. These condensates can sustain vortices, which have been envisioned to be used in applications for interferometry, as for example gyroscopy with counter-rotating vortex superpositions \cite{interferometry}, quantum information, such as coherent superpositions of arbitrary winding numbers \cite{QI1} or entangled vortex states \cite{QI2}, and as a way to study the behavior of random polynomial roots \cite{poly}. Different techniques have been proposed and experimentally reported to generate angular momentum with single atoms and BECs, such as stirring with a laser beam \cite{stirring}, phase imprinting \cite{phaseimp}, transfer of orbital angular momentum from optical states \cite{OAM, AMSA}, rotating traps \cite{rotating}, turbulence \cite{turbulence}, dynamical instabilities \cite{instabilities} or merging multiple trapped BECs \cite{merging}.

At the same time, adiabatic techniques to control the external degrees of freedom of massive particles have been developed \cite{TLAO}, based on the spatial analogue of the Stimulated Raman Adiabatic Passage technique \cite{STIRAP}. For the centre-of-mass degree of freedom, this is usually realized by considering a triple well configuration and assuming that only  a single state in each trap contributes to the dynamics. Up to now, all proposals in which the spatial adiabatic passage technique has been discussed have been effectively one-dimensional (1D): the traps are arranged in a linear geometry and a single particle in one of the outer traps is coherently transferred to the other outermost trap with very high fidelity. Significant work has been done for this process by discussing efficiency and robustness for single atoms \cite{TLAO, SAP}, electrons \cite{greentree}, atomic vortices \cite{Suzanne}, holes \cite{Holes} and BECs \cite{BECs}. Proposals for spatial adiabatic passage for cold atoms propagating in systems of three coupled waveguides have also been discussed using effective 1D models \cite{dip_waveguides}.
Recently, spatial adiabatic passage for light propagating in a system of three coupled optical waveguides \cite{light} has been experimentally reported.

In this work, we go beyond those well understood 1D spatial adiabatic passage systems and focus onto the possibilities offered by their extension to two dimensions (2D). The inclusion of this new degree of freedom allows considering novel scenarios in which all traps can be tunnel-coupled simultaneously. Contrarily to the well-discussed one-dimensional models for spatial adiabatic passage, the two-dimensional scenario has no fully equivalent model in quantum optics, where the STIRAP technique involves only two laser couplings of adjacent transitions of a three-level atomic system. Furthermore, understanding the fundamentals of 2D spatial adiabatic passage opens the possibility to study physical scenarios that require a full 2D description, such as the implementation of new interferometric schemes \cite{menchon-enrich_single-atom_2014}, which take advantage of the existing level crossings, or the generation of angular momentum carrying states.
Angular momentum is an inherent 2D quantity, which can only be created in systems in which rotational symmetry is broken. We demonstrate that, by applying a spatial adiabatic passage sequence in a system of three traps with broken spatial symmetry, a single particle can be completely transferred from the ground vibrational state of the initial trap to the two degenerate first excited states of the final trap. Depending on the total time of the process, this can generate angular momentum with values oscillating between $\pm\hbar$. Furthermore, the process is robust since both, the complete transfer and the generation of angular momentum, occur within a broad range of parameter values. We model the generation of angular momentum by using the asymptotic states of the individual traps and the results are confirmed with a numerical integration of the full 2D Schr\"odinger equation. Note that such a two-dimensional process constitutes an alternative method to standard techniques for the generation of angular momentum in ultracold atoms \cite{stirring,phaseimp,OAM, AMSA,rotating,turbulence,instabilities,merging}.

\section{Physical system}
\begin{figure}[!ht]
\centerline{
\includegraphics[width=\linewidth]{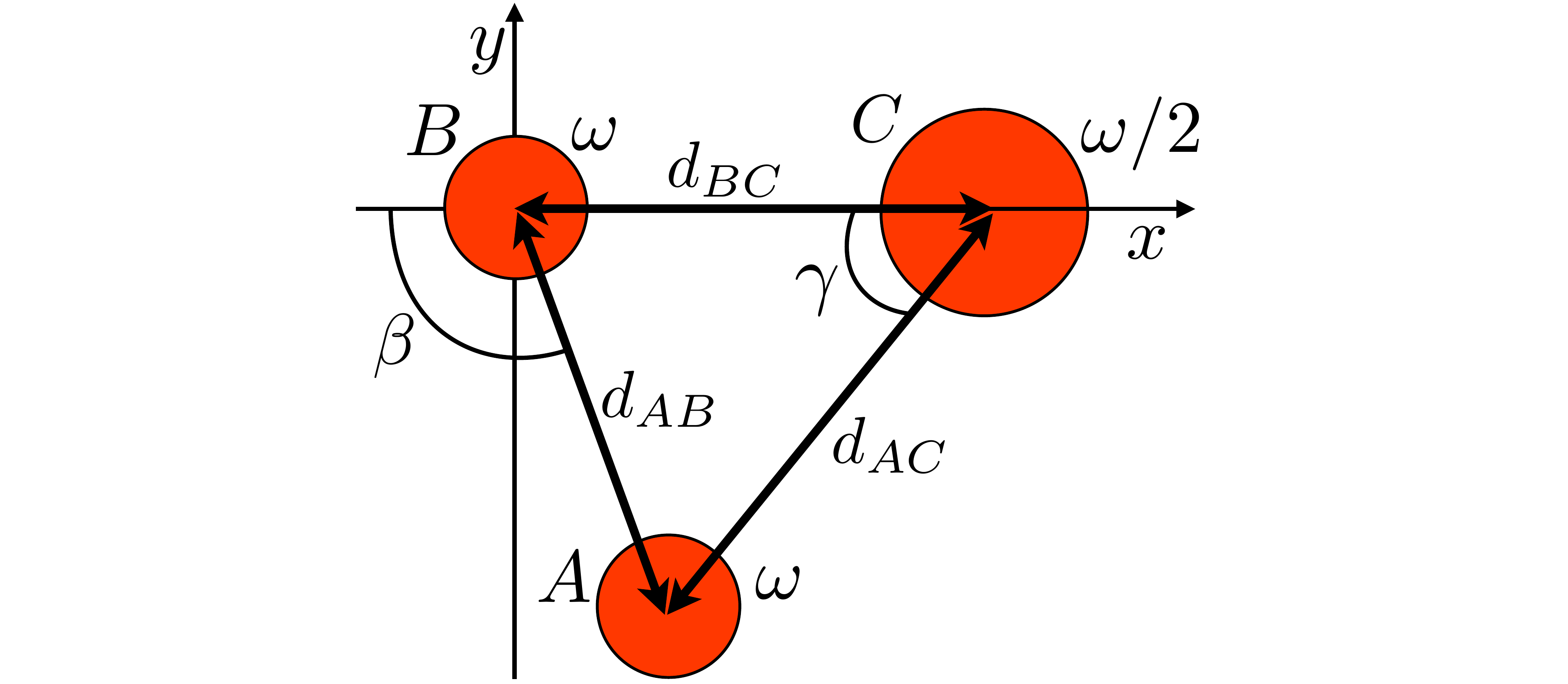}}
\caption{(Color online) Schematic representation of the system of three harmonic traps with broken spatial symmetry. The traps $A$ and $B$ have a trapping frequency $\omega$ and trap $C$ has a trapping frequency $\omega/2$.}
\label{fig1}
\end{figure}
We consider a system consisting of three 2D harmonic potentials (labeled $A$, $B$ and $C$), see \fref{fig1}, where initially a single particle is located in the vibrational ground state of trap $A$. The trapping frequencies of $A$ and $B$ are equal ($\omega_A=\omega_B=\omega$) and the one for $C$ is half that value ($\omega_C=\omega/2$), so that a resonance exists between the ground levels of traps $A$ and $B$ and the first excited level of trap $C$. Since the first excited energy level of trap $C$ is double degenerate, it supports an angular momentum carrying state through a superposition of the two energy eigenstates $\psi^{C}_{1,0}(x,y)$ and $\psi^{C}_{0,1}(x,y)$ in the chosen $x$--$y$ reference frame. In particular, maximum angular momentum, $\langle L_z\rangle=\pm\hbar$, occurs when the two degenerate states are (i) equally populated and (ii) have a phase difference of $\pi/2$, i.e.:
\begin{equation}
\psi^C_{n=1,l=\pm 1}(r,\theta)=\frac{1}{\sqrt{2}}\left[\psi^C_{1,0}(x,y)\pm i\psi^C_{0,1}(x,y)\right].
\label{max}
\end{equation}
Here the $\psi^C_{n=1,l=\pm 1}(r,\theta)$ are the eigenfunctions of the first excited states of the $C$ trap in polar coordinates, with $n$ and $l$ being the principal quantum number and the $z$ component of the angular momentum, respectively.

The first thing to note is that an effective 1D configuration in which all traps are arranged in a straight line along the $x$ axis can only lead to population transfer to the state $\psi^C_{1,0}(x,y)$ and not to the generation of angular momentum. To populate both of the degenerate states it is necessary to break the symmetry of the linear configuration and here we do this by considering geometries in which the $A$ trap is rotated around the $B$ trap and forms an angle $\beta$ with respect to the $x$ axis, as can be seen in \fref{fig1}.  The positions of the trap centers are $x_A=-d_{AB}\cos\beta$,  $y_A=-d_{AB}\sin\beta$,  $x_B=y_B=0$,  $x_C=d_{BC}$ and  $y_C=0$, where $d_{AB}$ and $d_{BC}$ are the distances between the $A$ and $B$ and the $B$ and $C$ traps, respectively. In the $x$--$y$ reference frame, the ground states in the $A$ and $B$ traps can be expressed as $\psi^{A}_{0,0}=\phi^\omega_0(x+d_{AB}\cos\beta)\phi^\omega_0(y+d_{AB}\sin\beta)$
and $\psi^{B}_{0,0}=\phi^\omega_0(x)\phi^\omega_0(y)$, respectively. For the $C$ trap with $\omega/2$ we consider the eigenfunctions $\psi^{C}_{1,0}=\phi^{\omega/2}_1(x-d_{BC})\phi^{\omega/2}_0(y)$ and $\psi^{C}_{0,1}=\phi^{\omega/2}_0(x-d_{BC})\phi^{\omega/2}_1(y)$. Here $\phi^{\tilde{\omega}}_n$ are the single-particle eigenfunctions for the $n$-th vibrational state of the 1D quantum harmonic oscillator with trapping frequency $\tilde{\omega}$.

The breaking of symmetry introduces two new effects with respect to effective 1D configurations: (i) the traps $A$ and $C$ can get close enough to allow direct tunnel coupling and (ii)  there is no longer a preferred direction along which one of the states in $C$ can line up. Thus, since the position of the $A$ trap with respect to the $C$ trap forms an angle $\gamma$ with the $x$ axis, there is a population transfer between the state $\psi^{A}_{0,0}$ and $\psi^{C}_{1,0}$ with coupling rate $J^{AC}_{1,0}$, but also between $\psi^{A}_{0,0}$ and $\psi^{C}_{0,1}$ with coupling rate $J^{AC}_{0,1}$. Therefore, both first excited states of the $C$ trap become involved in the dynamics of the system, which in the basis $\{\psi^{A}_{0,0}, \psi^{B}_{0,0}, \psi^{C}_{1,0},\psi^{C}_{0,1}\}$ can be described by the $4\times 4$ Hamiltonian
\begin{equation}
H =\hbar\begin{pmatrix}
0 & -J^{AB} & -J^{AC}_{1,0} & -J^{AC}_{0,1}\\
-J^{AB} & 0 & -J^{BC} & 0\\
-J^{AC}_{1,0} & -J^{BC} & 0 & 0\\
-J^{AC}_{0,1} & 0 & 0 & 0
\end{pmatrix}.
\label{ham}
\end{equation}
The coupling rates depend on the trap separation and can be calculated analytically due to the harmonicity of the potentials \cite{TLAO}. Additionally, the coupling rates between the ground state in $A$ and the first excited states in $C$ depend on the angle $\gamma$ as $J^{AC}_{1,0}=J_{\omega,\omega/2}\cos\gamma$ and $J^{AC}_{0,1}=J_{\omega,\omega/2}\sin\gamma$, where $J_{\omega,\omega/2}$ is the coupling rate between the ground state of a trap with trapping frequency $\omega$ and a resonant first excited state of a trap with trapping frequency $\omega/2$. It is straightforward to check that Hamiltonian (2) possesses four non-degenerate eigenvalues, except for 
$J^{BC}/\sqrt{2}=J^{AB}=J^{AC}_{1,0}=J^{AC}_{0,1}$ when two of them become degenerate. 

\section{Generation of angular momentum carrying states}
\begin{figure}[tbh]
\centerline{
\includegraphics[width=\linewidth]{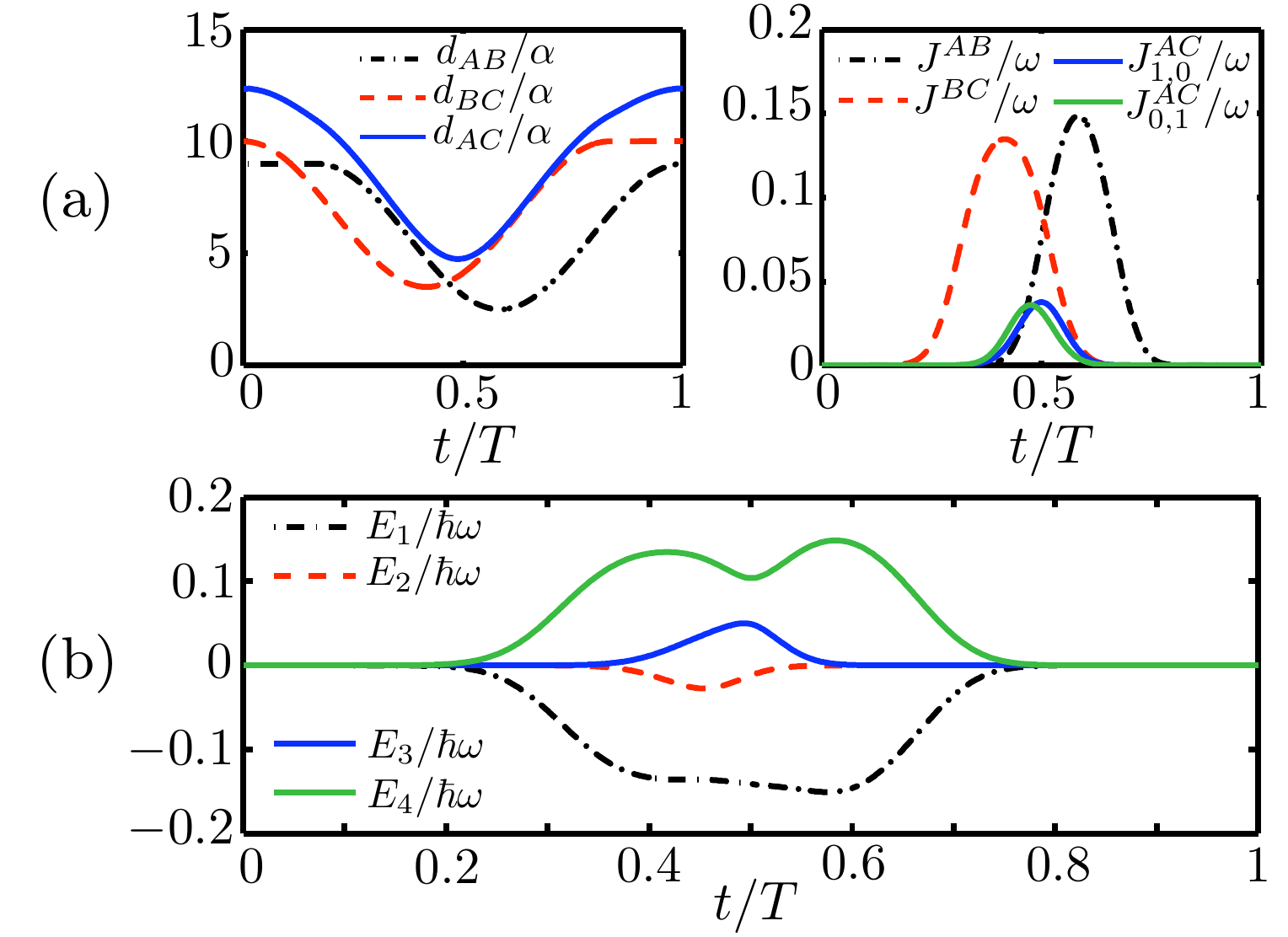}}
\caption{(Color online) (a) Temporal evolution of the distances between traps $d_{AB}$, $d_{BC}$ and $d_{AC}$ (left panel), and the couplings $J^{AB}$, $J^{BC}$, $J^{AC}_{1,0}$ and $J^{AC}_{0,1}$ (right panel), during the spatial adiabatic passage process. (b) Energy eigenvalues as a function of time. The parameter values are: $\beta=0.55\pi$, $\delta=0.2T$, and $d_{BC}$ and $d_{AB}$ with values between $10\alpha$ and $3.5\alpha$, and $9\alpha$ and $2.5\alpha$, respectively, where $\alpha=\sqrt{\hbar/(m\omega)}$, $m$ is the mass of the particle and $T$ is the total time of the process.}
\label{fig2}
\end{figure}

The generation of angular momentum occurs along with the transfer of the particle from the $A$ to the $C$ trap through a spatial adiabatic passage process \cite{TLAO,SAP}. This corresponds to a counterintuitive temporal sequence of the couplings, i.e. with the particle initially located in $A$ and the position of $B$ being fixed, the approach and separation sequence of the $C$ trap towards the $B$ trap along the $x$ axis is initiated a time $\delta$ before the $A$ trap approaches and separates from the $B$ trap, keeping the angle $\beta$ constant. In the following, we will analyze this process in terms of the overall energy eigenvalues and eigenstates of the system by diagonalizing the Hamiltonian in \eref{ham}. To approach and separate the traps, the evolution of the distances $d_{BC}$ and $d_{AB}$ follows a cosine function evaluated between $0$ and $2\pi$, see \fref{fig2}(a) left. The right hand side panel of \fref{fig2}(a) shows the corresponding tunneling rates, and \fref{fig2}(b) displays the temporal evolution of all four energy eigenvalues of the Hamiltonian. Note that for the chosen parameters no level crossing occurs. 

\begin{figure}[htb]
\centerline{
\includegraphics[width=1\linewidth]{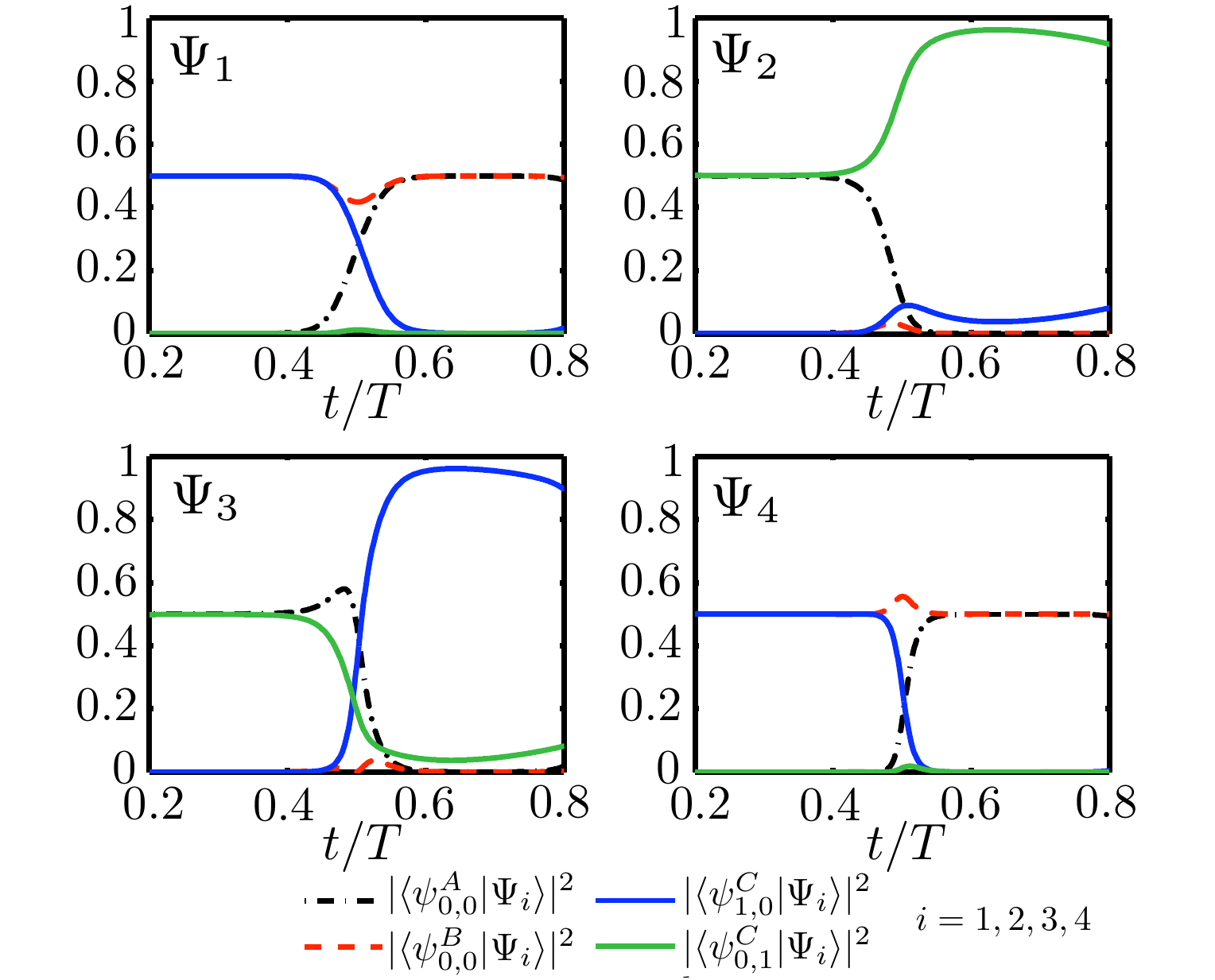}}
\caption{(Color online) Temporal evolution of the population of each asymptotic level of the traps ($\psi^A_{0,0}$, $\psi^B_{0,0}$, $\psi^{C}_{1,0}$ and $\psi^{C}_{0,1}$) for the four eigenstates of the system ($\Psi_1$, $\Psi_2$, $\Psi_3$ and $\Psi_4$). The parameter values are as in \fref{fig2}.}
\label{fig3}
\end{figure}

The population of each asymptotic level of the traps for the four energy eigenstates of the system is shown in \fref{fig3}. Since initially the particle is in the $A$ trap, the eigenfunction of the system at $t=0$ can be written as a superposition of the eigenstates $\Psi_2$ and $\Psi_3$ as
\begin{equation}
   \psi(t=0)=\frac{1}{\sqrt{2}}\left[\Psi_2(t=0)+\Psi_3(t=0)\right]=\psi^A_{0,0},
\end{equation}
where $\Psi_2(t=0)=\left(\psi^A_{0,0}+\psi^C_{0,1}\right)/\sqrt{2}$ and $\Psi_3(t=0)=\left(\psi^A_{0,0}-\psi^C_{0,1}\right)/\sqrt{2}$. If the process is adiabatic and level crossings are absent, this superposition of eigenstates is followed all through the process, leading to a final state of the form
\begin{align}
 \psi(t=T)=\frac{1}{\sqrt{2}}\Biggl[\Psi_2(t=T)\exp\left(-\frac{i}{\hbar} \int_0^TE_2(t)dt\right)+\notag \\
 \Psi_3(t=T)\exp\left(-\frac{i}{\hbar}\int_0^TE_3(t)dt\right)\Biggl].
\label{sup}
\end{align}

As it can be seen from \fref{fig3}, at the end of the process the eigenstates $\Psi_2$ and $\Psi_3$ only involve the asymptotic states of the $C$ trap: $\Psi_2(t=T)=a\psi^{C}_{0,1}-b\psi^{C}_{1,0}$ and $\Psi_3(t=T)=b\psi^{C}_{0,1}+a\psi^{C}_{1,0}$, where $|a|^2+|b|^2=1$. This means that by following the two eigenstates $\Psi_2$ and $\Psi_3$, a complete transfer of population from the initial trap $A$ to the final trap $C$ is achieved. Noting that the superpositions $(a\psi^{C}_{0,1}-b\psi^{C}_{1,0})$ and $(b\psi^{C}_{0,1}+a\psi^{C}_{1,0})$ are also two asymptotic eigenstates of the $C$ trap in a reference frame rotated with respect to $\psi^{C}_{1,0}$ and $\psi^{C}_{0,1}$, we call them $\psi'^{C}_{0,1}=\Psi_2(t=T)$ and $\psi'^{C}_{1,0}=\Psi_3(t=T)$. It is then easy to see from \eref{sup} that  $\psi'^{C}_{0,1}$ and $\psi'^{C}_{1,0}$ are equally populated and that  the phase difference between them is given by
\begin{equation}
\varphi(t=T)=\frac{1}{\hbar}\int_0^T(E_3(t)-E_2(t))dt.
\label{desf}
\end{equation}
From \fref{fig2}(b) it can be seen that the phase difference will be directly proportional to $T$, since the energy difference between the energy eigenvalues $E_3$ and $E_2$ follows the same pattern, independent of the total time of the process. Rewriting \eref{sup} in polar coordinates it is possible to show that the expected value of the angular momentum is a function of the phase difference $\varphi$
\begin{equation}
\langle L_z(T)\rangle=\hbar\sin[\varphi(T)].
\label{l}
\end{equation}

Therefore, the generated angular momentum $\langle L_z\rangle$ will follow a sinusoidal curve as a function of $T$ with a maximum at $\varphi=(n+1/2)\pi$ for $n\in\mathbb{N}$. We have checked that the above process works for all angles $0< \beta<\beta_t$,  which is the parameter range in which the energy eigenspectrum and eigenfunctions are similar to the ones shown in \fref{fig2} and \ref{fig3}. The only significative variation is that the energy difference between $E_3$ and $E_2$ is smaller for smaller angles, leading to longer oscillation periods of the angular momentum as a function of $T$. Around $\beta_{t}\approx0.625\pi$, the energy eigenstates $\Psi_3$ and $\Psi_4$ become almost degenerate at one point during the evolution, which limits the possibility to follow the superposition of the states $\Psi_2$ and $\Psi_3$ adiabatically. As discussed earlier, this particular time corresponds to the instant at which $J^{BC}/\sqrt{2}\approx J^{AB}\approx J^{AC}_{1,0}\approx J^{AC}_{0,1}$ and $\beta_{t}$ therefore represents the angle up to which both, a complete transfer and the generation of angular momentum, work efficiently. 

\section{Numerical simulations}
Although above we have used the asymptotic states of the individual traps to describe the dynamics of the system, the full dynamics is governed by the 2D Schr\"odinger equation
\begin{equation}
i \hbar\frac{\partial }{\partial t}\psi (x,y)=\left[
- \frac{\hbar^2}{2m}
\nabla^2 
+ {V}\left( x,y \right)\right ] \psi (x,y), \label{Eq_Schrodinger_2D}
\end{equation}
where $\nabla^2$ is the 2D Laplace operator and $V(x,y)$ is the trapping potential, which we assume to be constructed from truncated harmonic oscillator potentials
\begin{equation}
   V( x,y) =\min_{i=A,B,C} \left\{\frac{1}{2}m\omega_i^2\left[(x-x_i)^2+(y-y_i)^2 \right] \right\}.
   \label{trun_pot}
\end{equation}
Here $(x_i,y_i)$ with $i=A,B,C$ are the positions of the individual trap centers, $\omega_A=\omega_B=\omega$ and $\omega_C=\omega/2$. To establish the validity of our model above, we present in the following the numerical solution of \eref{Eq_Schrodinger_2D}, with the trapping potential \eref{trun_pot} in the regime where $\beta<\beta_t$.

\begin{figure}[!ht]
\centerline{
\includegraphics[width=1\linewidth]{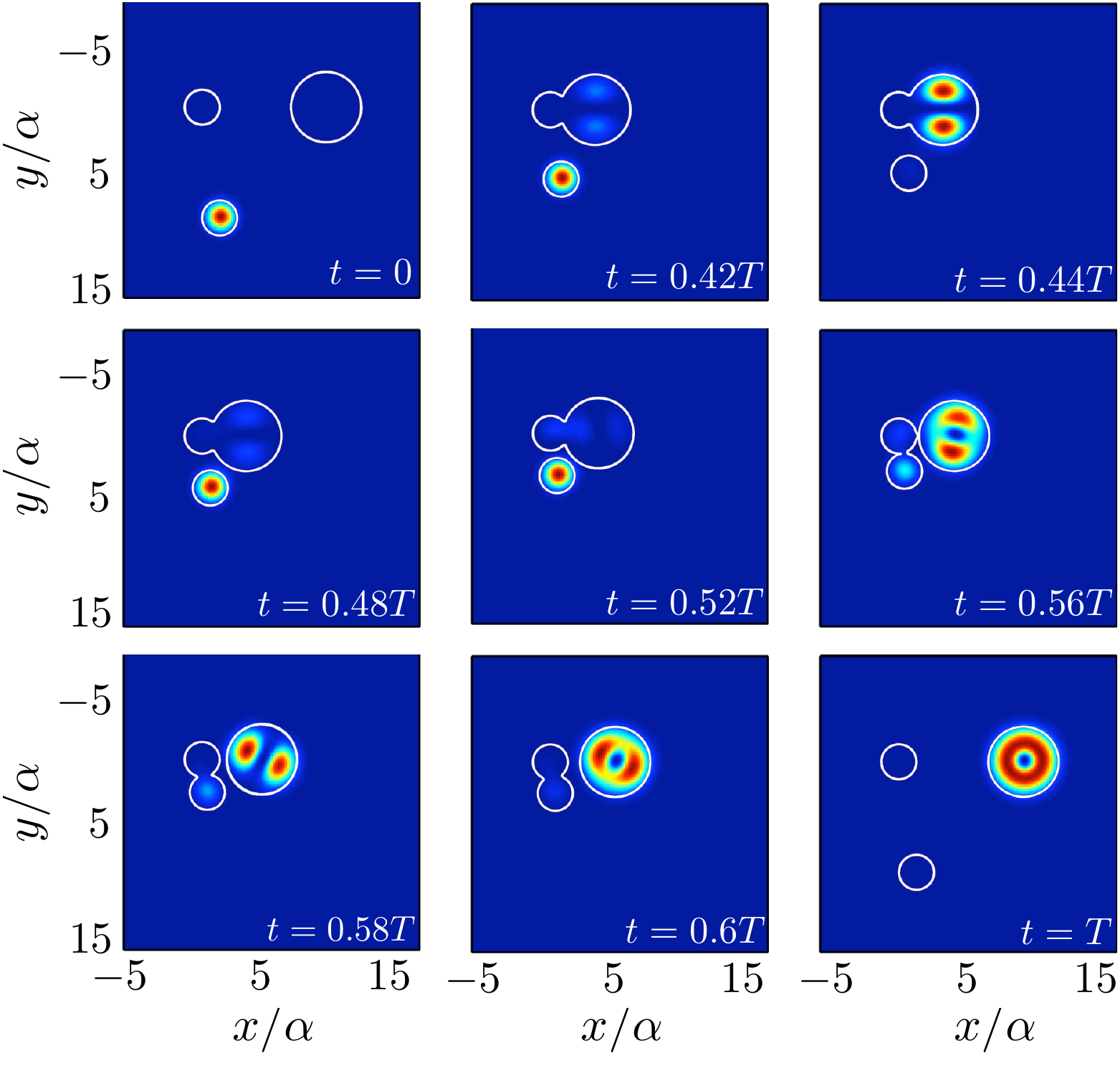}}
\caption{(Color online) Temporal evolution of the population distribution of the single particle in the system of three traps with broken spatial symmetry for $T=5183\omega^{-1}$ and the parameter values as in \fref{fig2}. The adiabatic following of the superposition of $\Psi_2(t)$ and $\Psi_3(t)$ can be observed by comparing with \fref{fig3}: from $t=0.42T$ to $t=0.48T$ oscillations corresponding to the phase difference between $\Psi_2(t<0.5T)=(\psi^A_{0,0}+\psi^C_{0,1})/\sqrt{2}$ and $\Psi_3(t<0.5T)=(\psi^A_{0,0}-\psi^C_{0,1})/\sqrt{2}$ are shown; for $t=0.58T$ and $t=0.6T$ oscillations due to the phase difference between $\Psi_2(t>0.5T)= \psi'^{C}_{0,1}$ and $\Psi_3(t>0.5T)=\psi'^{C}_{1,0}$ are also observed.}
\label{fig4}
\end{figure}

\begin{figure}[!ht]
\centerline{
\includegraphics[width=1\linewidth]{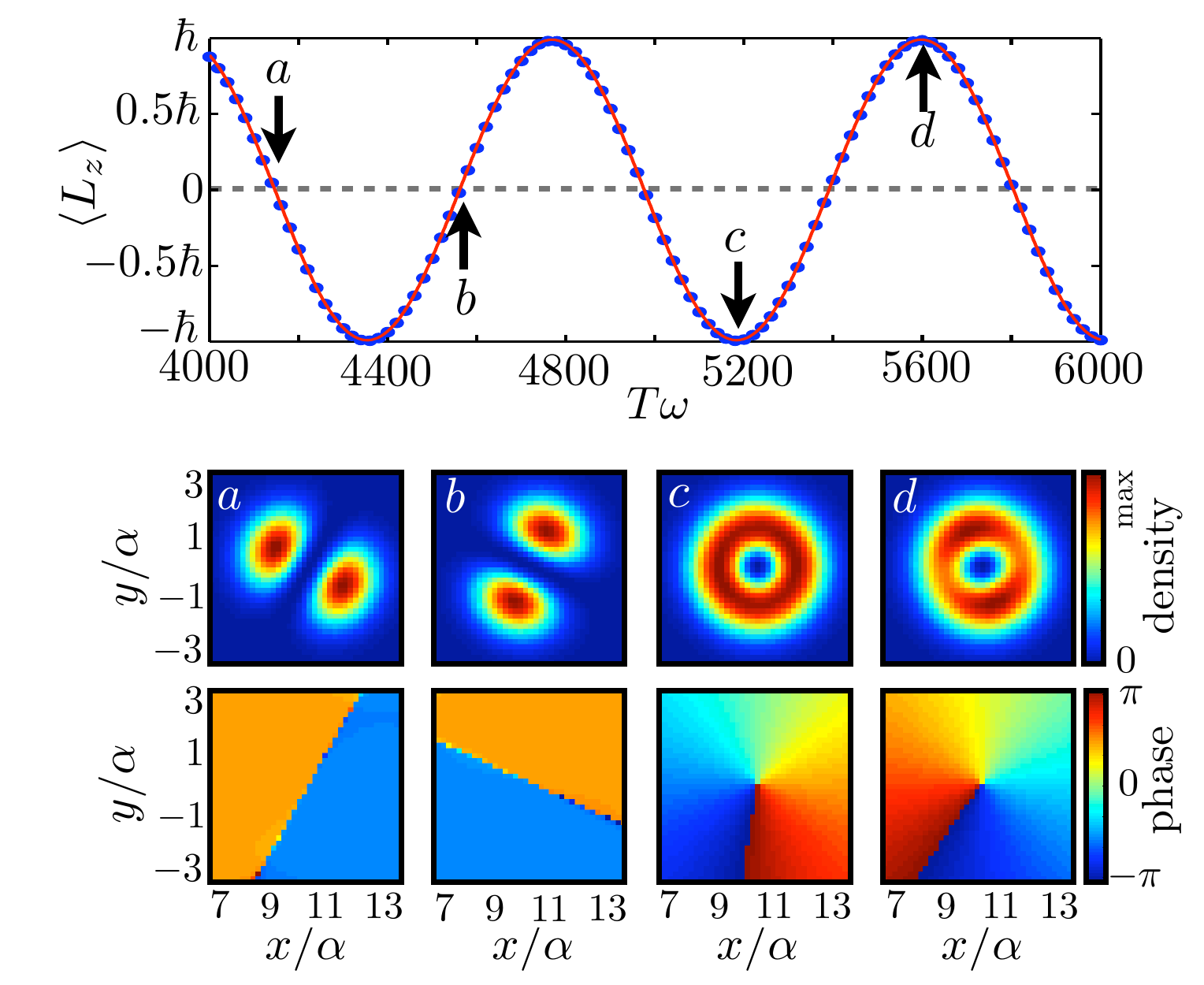}}
\caption{(Color online) Generated angular momentum as a function of the total time of the process T (upper plot). Final states in the $C$ trap and their phases for four different final total times $a$, $b$, $c$ and $d$ (lower plots). Parameter values as in \fref{fig2}.}
\label{fig5}
\end{figure}

The population distribution at different times is shown in \fref{fig4} for a process of total time $T=5183\omega^{-1}$. One can see that a single particle is completely transferred from the $A$ trap to the $C$ trap, where a state with maximum angular momentum, $\langle L_z\rangle=-\hbar$, is created, which corresponds to the adiabatic following of the eigenstates $\Psi_2$ and $\Psi_3$ (see also point {\sl c} in \fref{fig5}).

In agreement with \eref{l}, \fref{fig5} shows the sinusoidal behaviour of the angular momentum generated in the $C$ trap as a function of $T$, with maximum values of $\pm\hbar$. We have numerically checked that, as long as the process is performed adiabatically, both the generation of angular momentum and the complete population transfer into the $C$ trap work efficiently over a broad range of parameter values, including different distances between the traps and angles ranging from very small values of $\beta$ up to $\beta_t\approx0.625\pi$. Thus, the process is very robust and highly versatile.

Let us finally briefly comment on a possible implementation of the above scheme using current technology. Optical microtraps, where atoms are trapped in the focal point of a red-detuned laser beam, can be used to create versatile potentials for single atoms and the presented triple well system can, for example, be build using holographic techniques with programmable spatial light modulators \cite{Bergamini:04}. This setup allows for the storage of a single atom per trap and the dynamical control of the trapping potentials, especially on time-scales required by the adiabaticity condition. Note that the process we describe can also be realized for light in three-dimensional, laser-written optical waveguide arrays \cite{Marshall:09}.

\section{Conclusions}
In conclusion, we have extended the spatial adiabatic passage technique to the 2D case, demonstrating that angular momentum can be successfully generated by breaking the symmetry of the coupling sequence in a system of three harmonic traps of different trapping frequencies. Starting with a single particle in the ground state of a harmonic trap with frequency $\omega$, it can be fully transferred to the degenerate first excited states of a final harmonic trap of frequency $\omega/2$ by following adiabatically a superposition of two energy eigenstates of the system. The energy difference between these two eigenstates results in a phase difference between the equally populated excited states of the final harmonic trap, which leads to the generation of angular momentum. The obtained values oscillate between $\pm\hbar$ and depend on the total time of the process. We have modelled this process by using the asymptotic levels of the three harmonic traps and checked the results against the numerical solution of the full two-dimensional Schr\"{o}dinger equation for a broad range of parameter values.  Our work shows that adiabatic techniques for centre-of-mass states hold significant potential for new processes, that have no direct equivalent in, for example, the control of internal degrees of freedom.

\section*{Acknowledgments}
The authors gratefully acknowledge financial support through the Spanish MICINN contract FIS2011-23719, the Catalan Government contract SGR2009-00347. R. M.-E. acknowledges financial support from AP2008-01276 (MECD) and also Albert Benseny for fruitful discussions. T.B. acknowledges support by Science Foundation Ireland under project number 10/IN.1/I2979.

\end{document}